\documentclass[aps,prapplied,reprint,preprintnumbers,longbibliography]{revtex4-2} %superscriptaddress
\usepackage{hyperref}
\usepackage{amsmath,latexsym}
\usepackage{graphicx}
\usepackage{siunitx}
\usepackage{textcomp}
\usepackage{wasysym}
\usepackage{amssymb}
\usepackage{MnSymbol}
\usepackage{oplotsymbl}
\usepackage{physics}
\usepackage{array}
\usepackage{tikz}

\DeclareSIUnit{\molar}{M}
\DeclareSIUnit{\rev}{rev}
\DeclareSIUnit{\wtpercent}{wt\%}
\begin{document}
\title{Spreading of Low-viscosity Ink Filaments \\ Driven by Bath Viscoelasticity in Embedded Printing}
\date{\today}
\author{Jae Hyung Cho}
\affiliation{Department of Mechanical Engineering, University of California, Santa Barbara, California 93106, USA}
\author{Emilie Dressaire}
\email[To whom correspondence should be addressed. ]{dressaire@ucsb.edu}
\affiliation{Department of Mechanical Engineering, University of California, Santa Barbara, California 93106, USA}

\begin{abstract}
Inks deposited in conventional direct ink writing need to be able to support their own weight and that of the upper layers with minimal deformation to preserve the structural integrity of the three-dimensional (3D) printed parts. This constraint limits the range of usable inks to high-viscosity materials. Embedded printing enables the use of much softer inks by depositing the materials in a bath of another fluid that provides external support, thus diversifying the types of 3D printable structures. The interactions between the ink and bath fluids, however, give rise to a unique type of defect: spreading of the dispensed ink behind the moving nozzle. By printing horizontal threads made of dyed water in baths of Carbopol suspensions, we demonstrate that the spreading can be attributed to the pressure field generated in the viscous bath by the relative motion of the nozzle. As the pressure gradient increases with the viscosity of the bath fluid while the viscosity of the ink resists the flow, a larger bath-to-ink viscosity ratio results in more spreading for low-concentration Carbopol baths. For high-concentration, yield-stress-fluid baths, we find that the steady-state viscosity alone cannot account for the spreading, as the elastic stress becomes comparable to the viscous stress and the bath fluid around the dispensed ink undergoes fluidization and resolidification. By parameterizing the transient rheology of the high-concentration Carbopol suspensions using a simple viscoelastic model, we suggest that the ink spreading is exacerbated by the elasticity but is mitigated by the yield stress as long as the yield stress is low enough to allow steady injection of the ink. These results help illuminate the link between the bath rheology and the printing quality in embedded 3D printing.
\end{abstract}

\maketitle

%Introduction
\section{Introduction}

The geometry of three-dimensional (3D) structures built via conventional direct ink writing is inherently constrained by the effect of gravity; the ink has to be dispensed onto previously deposited filaments, and these filaments have to be sufficiently stiff to support the parts that lie above themselves \cite{Saadi2022,Liu2023}. Embedded 3D printing opens up a range of printable geometries by dispensing the ink in a bath of viscous fluid. The bath fluid holds the dispensed ink filament in place against the gravity, thus having the ink ``embedded'' in its matrix \cite{Wu2011,Muth2014,Bhattacharjee2015,LeBlanc2016,OBryan2017,Highley2019,Grosskopf2018,Lee2019,Noor2019,Cai2019,Uchida2019,Jin2019,Ning2020,Friedrich2020,Xu2020,Cooke2021,Prendergast2022,Wu2022,Friedrich2022,Friedrich2022a,Becker2023,Arun2023,Duraivel2023,Trikalitis2023}. Removing the need to stack up printed filaments enables fabrication of delicate structures, such as overhangs, and renders runny ink fluids that cannot otherwise retain their own shapes compatible with direct ink writing. Use of less stiff, low-viscosity fluids as inks may be desirable especially in bioprinting. Lower ink viscosity decreases the injection pressure, which induces a lower stress field that cells in the ink would have to withstand during the flow, hence improving the cell viability within the printed structure \cite{Cooke2021,Becker2023}. \par

\begin{figure}[t]
\setlength{\abovecaptionskip}{7pt}
\hspace*{0cm}\includegraphics[scale=0.5]{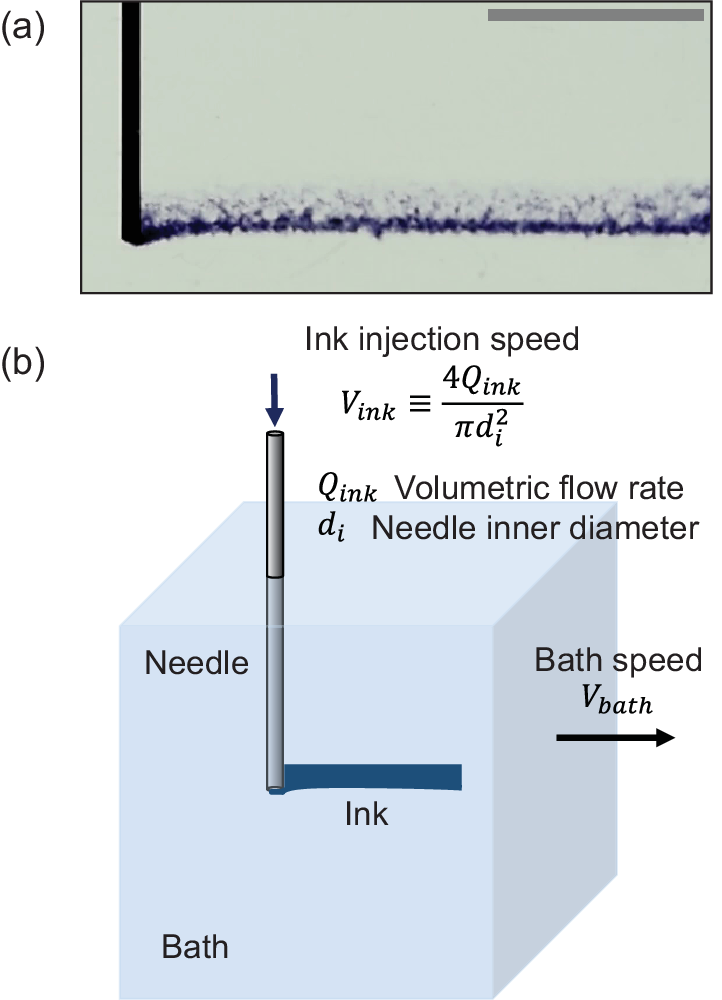}
\caption{\label{ink_spreading} (a) Vertical spreading of ink observed in embedded printing. As the nozzle translates to the left, the ink in the deposited horizontal thread partially spreads upward right behind the needle, forming an undesirable thin layer. Scale bar represents $10\;\si{\milli\meter}$. (b) Schematic of the experimental setup.}
\end{figure}

The presence of the bath fluid in embedded printing, however, gives rise to a unique type of structural defect -- ink spreading behind the nozzle. Such spreading of the ink can be readily observed by injecting a low-viscosity fluid into a bath of another fluid while the nozzle translates in a direction perpendicular to its orientation, as shown in Fig.~\ref{ink_spreading}(a). For more viscous inks, similar phenomena, such as the vertical displacement of printed filaments \cite{LeBlanc2016,Cai2019,Uchida2019,Friedrich2022a,Prendergast2022} or the vertical elongation of the filament cross section \cite{Jin2019,Friedrich2021,Friedrich2022,Friedrich2022a,Arun2023}, have been reported. The origin of the upward driving force, however, remains largely unexplained. \par

The upward ink flow behind the nozzle is driven by a nontrivial flow of the surrounding bath fluid, often compounded by its complex rheological properties. For a Newtonian fluid, the momentum equation of the flow past an infinite cylinder in the viscosity-dominated regime cannot be solved without including an advective inertial term \cite{Kundu2016}. The flow of a viscoelastic fluid around a cylinder features pressure and velocity fields highly sensitive to the relative magnitudes of the viscous and the elastic forces \cite{Kenney2013,Haward2018,Li2019}. The flow of a yield stress fluid that exhibits solidlike deformations below a stress threshold displays pronounced spatial heterogeneities, as the fluid is locally fluidized around the cylinder \cite{Putz2008,Tokpavi2008,Tokpavi2009}. Embedded printing involves multiple factors that further complicate the flow, such as the end effect at the tip of the nozzle \cite{Li2019}, physicochemical interactions between the ink and the bath fluids \cite{Friedrich2020,Friedrich2021,Becker2023}, elastoplastic deformations of the unyielded regions of the yield stress fluids \cite{Piau2007}, and flow-history dependence of the rheological parameters of thixotropic fluids \cite{Dinkgreve2018,Younes2020}. Given this complexity, one possible approach is to investigate how a selected group of key rheological material functions affects the phenomenon in a relatively simple model system. \par

In this work, we explore how the viscoelasticity of the bath fluid governs the ink spreading by printing straight, horizontal filaments of dyed water in baths of aqueous Carbopol suspensions. Extensively used as bath fluids because of their optical transparency and rheological tunability, Carbopol suspensions are mixtures of water and polyacrylic-acid-based microgel particles that can serve as a model shear thinning fluid at low concentrations and a model yield stress fluid at high concentrations \cite{Gutowski2012,Bhattacharjee2018,Dinkgreve2018,Jaworski2022,Oelschlaeger2022}. At low concentrations, the microgel particles are free to move around under shear, giving rise to shear-thinning behavior typical of suspensions whose particles interact hydrodynamically. At high concentrations, the microgel particles are jammed against each other, forming an elastic, space-spanning network of particles that induces a nonzero yield stress at the macroscopic level \cite{Gutowski2012,Oelschlaeger2022}. Far less viscous than typical ink materials, water as the ink can effectively spread due to the force applied by the Carbopol suspension in either concentration regime, thus enabling us to readily quantify the defect behind the nozzle. Moreover, the Newtonian nature of water allows itself to be fully characterized by a single material function -- a constant shear viscosity -- which facilitates identification of any correlations between the ink spreading and the bath fluid rheology. As the Carbopol suspension is also primarily composed of water, we neglect the effects of the interfacial tension \cite{Manglik2001,Boujlel2013,Mohammadigoushki2023}. \par

Our study identifies different spreading regimes depending on the concentration of the Carbopol suspensions. For Carbopol suspensions at low concentrations without yield stress, we find that the pressure drop across the nozzle due to the viscous flow of the bath fluid dictates the ink spreading. While a fluid element that travels around the nozzle undergoes a viscous pressure drop, an element that travels unperturbed below the tip of the nozzle is subject to a constant pressure. The resultant pressure gradient in the nozzle direction causes the length of spreading to scale as the square root of the bath-to-ink viscosity ratio. For Carbopol suspensions at high concentrations that exhibit nonzero yield stress, we suggest that the pressure field behind the nozzle is set by the elasticity as well as the viscosity of the bath fluid. The ink flow within the yielded region of the bath behind the translating nozzle is subject to stresses caused by both the viscous flow and the elastic deformation of the surrounding matrix. Plus, the bath fluid behind the nozzle constantly undergoes a fluid-to-solid transition as the microgel particles arrange themselves to recover a force-bearing network \cite{Gutowski2012,Oelschlaeger2022}. Using a set of model parameters that describe the transient state, we indeed identify comparable scales of the pressure gradients induced by the viscosity and the elasticity. The spreading at the high concentrations, however, is substantially less severe than that predicted by the viscoelasticity alone, and even decreases with the concentration for the most concentrated baths. We speculate that the decrease in the area of fluidization around the nozzle due to higher yield stress \cite{Tokpavi2008,Grosskopf2018} suppresses the viscous ink flow for these concentrations. \par   

Although the results may seem to suggest that the ink spreading in embedded printing is best controlled by utilizing Carbopol suspensions at very low or very high concentrations, we demonstrate that other types of defects may arise at such extreme concentrations. At the lowest concentrations, the low viscosity of the bath fluid proves insufficient to hold the printed filaments in place, whereas at the highest concentrations, the high yield stress, coupled with the mechanical compliance of the ink injection system, gives rise to intermittent dispensing of the ink. By documenting the structural defects observed in embedding printing with a low-viscosity ink, our study provides a guideline for optimizing the material rheology and print speed as well as predicting the spread of the ink. \par

%Methods
\section{Experimental methods}

\begin{figure*}[t]
\setlength{\abovecaptionskip}{-30pt}
\hspace*{0cm}\includegraphics[scale=0.20]{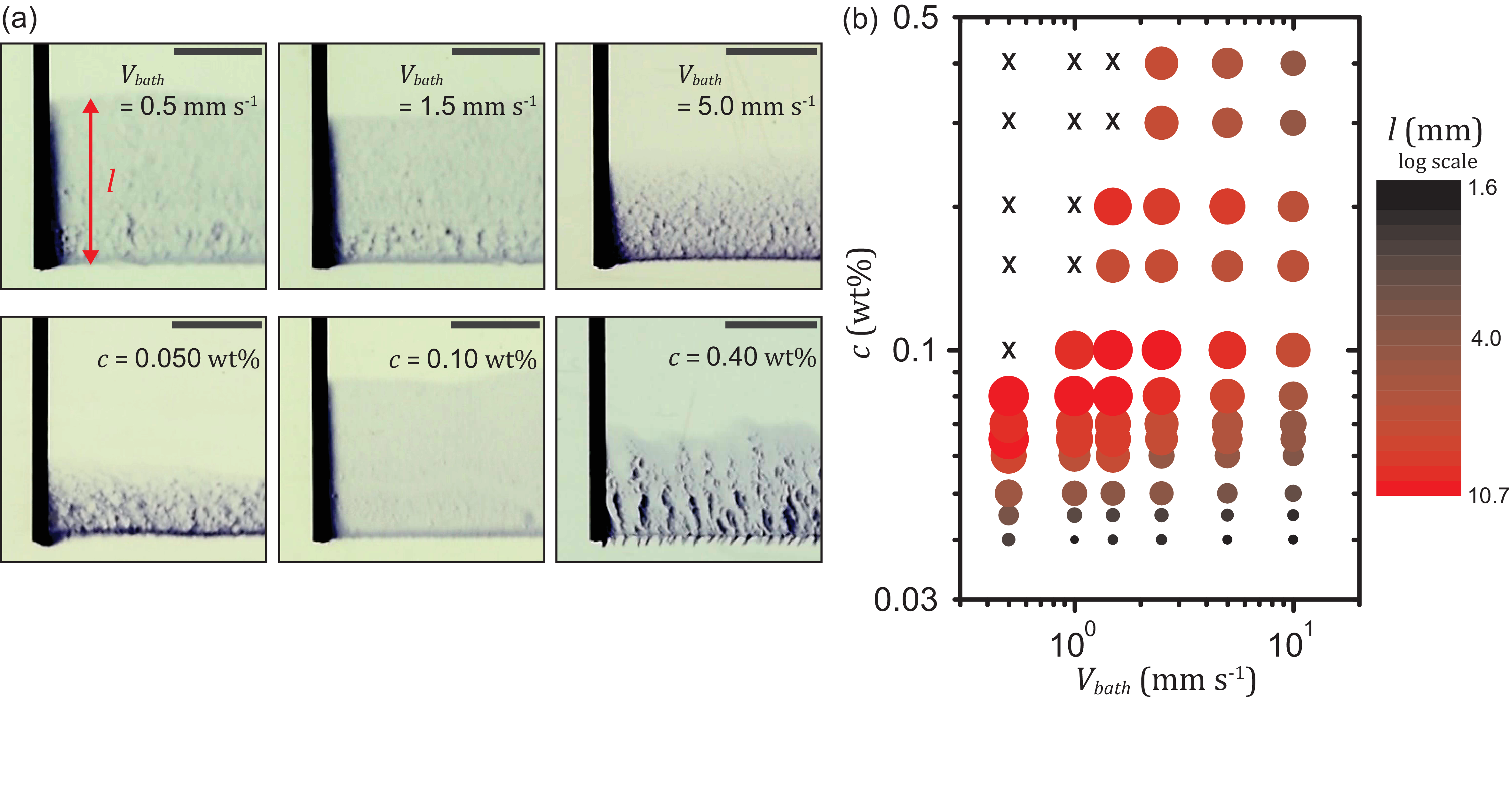}
\caption{\label{pics_heatmap} (a) Spreading of dyed water ink in Carbopol 940 baths for different bath speeds $V_{bath}$ at a concentration $c=0.070\,\si{\wtpercent}$ (top row) and for different concentrations $c$ at a bath speed $V_{bath}=5.0\,\si{\milli\meter\per\s}$ (bottom row). The penetration length of the ink in the needle direction is labeled as $l$. Each scale bar denotes $5\,\si{\milli\meter}$. (b) Penetration length $l$ for different Carbopol 940 concentrations $c$ and different bath speeds $V_{bath}$. The redder and the larger the circle, the higher the corresponding value of the penetration length. $l$ tends to decrease monotonically with $V_{bath}$, while it peaks at intermediate concentrations as a function of $c$. Symbol $\times$ indicates a highly unsteady ink flow, for which the mean penetration length cannot be reliably measured.}
\end{figure*}

\subsection{Material preparation}
To prepare a bath fluid, we add a specified amount of dry Carbopol powder (Carbopol 940 or Carbopol ETD 2050, Lubrizol) to $40\;\si{\milli\liter}$ of purified water (Milli-Q purification system, Millipore Sigma), and magnetically stir the sample for over $16\;\si{\hour}$ at room temperature to ensure complete mixing. We transfer the sample to a $40\times40\times40\;\si{\cubic\milli\meter}$ acrylic cube, and add $1\;\si{\molar}$  aqueous sodium hydroxide (NaOH) solution dropwise to neutralize the acidic mixture.  The sample is homogenized through mechanical stirring with a overhead mixer at $40\;\si{\rev\per\minute}$ for approximately $5\;\si{\minute}$. The rotation speed of the mixer is kept small to avoid trapping air bubbles in the Carbopol suspension. We let the sample sit at least for $24\;\si{\hour}$ before printing to minimize any possible effects of shear history formed during mixing and to further improve the homogeneity by the diffusion of the ions. \par

For the ink, we prepare a $0.1\;\si{\wtpercent}$ aqueous solution of a synthetic dye (Nigrosin, Sigma-Aldrich) by dissolving the dye in purified water under magnetic stirring for $5\;\si{\min}$ at room temperature. Given the low concentration of the solution, we assume that the viscosity of the ink is the same as that of pure water, $\eta_{ink}=1.0\,\si{\milli\pascal\second}$.\par

\subsection{Embedded printing}
We mount a cubic bath of Carbopol suspension onto a set of two perpendicularly connected ballscrew linear rails equipped with NEMA 17 stepper motors to move the bath in the vertical and a horizontal directions. In all our experiments, we move the bath only in the horizontal direction at a speed $V_{bath}=0.5 - 10.0\;\si{\milli\meter\per\s}$, while a 20-gauge stainless steel needle (inner diameter $d_i=0.603\;\si{\milli\meter}$ , outer diameter $d_o=0.908\;\si{\milli\meter}$) with a hydrophobic coating (LGN-GCC02, Liquid Glass) is held stationary in a vertical position, as shown in Fig.~\ref{ink_spreading}(b). The hydrophobic coating prevents any potential spreading driven by interfacial affinity between the needle and water. The needle is connected to a $10\;\si{\milli\liter}$ syringe on a syringe pump (KD Scientific) that dispenses the ink at a fixed volumetric flow rate $Q_{ink}$, which corresponds to the average ink injection speed $V_{ink}\equiv4Q_{ink}/\left(\pi {d_i}^2 \right)$. The ink injection speed $V_{ink}$ is kept the same as the bath translation speed $V_{bath}$, such that the volume of dispensed ink per unit length of the filament stays constant. The spreading occurs without delay right behind the needle on the timescale of the bath translation, orders of magnitude smaller than the ink diffusion timescale. \par

The printing process is recorded using a DSLR camera (Nikon D5300) equipped with a $105\;\si{\milli\meter}$ macro lens and an LED backlight panel (Phlox) for illumination. During the recording, the walls of the acrylic cube remain perpendicular to the optical path of the lens to minimize optical distortion due to refraction. We limit our analysis of the printing dynamics to its steady state by neglecting the transient behaviors at the beginning or the end of the bath translation. The vertical length of the spread ink layer in the steady state, measured from the tip of the needle, is denoted by the penetration length $l$, as displayed in Fig.~\ref{pics_heatmap}(a). \par 

\begin{figure*}[t]
\setlength{\abovecaptionskip}{-10pt}
\hspace*{0cm}\includegraphics[scale=0.5]{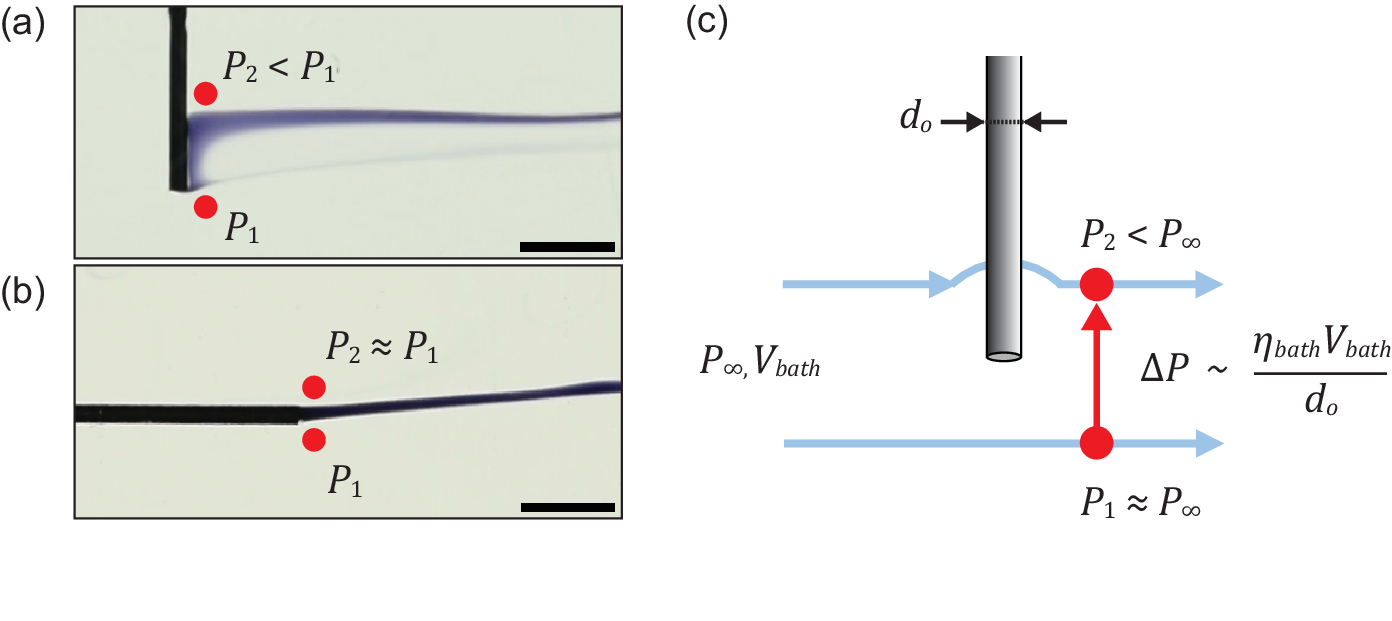}
\caption{\label{p_drop_N} Deposition of water threads in a glycerol bath that moves to the right at a constant speed $V_{bath}=5.0\,\si{\milli\meter\per\s}$ from (a) a vertically oriented needle and (b) a horizontally oriented needle. The upward ink flow behind the vertically oriented needle indicates a pressure gradient ($P_1>P_2$) in the direction parallel to the needle axis. Scale bar represents $5\;\si{\milli\meter}$. (c) Schematic of the streamlines above and below the needle tip. For a flow dominated by the viscous effects, the pressure difference in the needle direction ($\Delta P=P_1-P_2$) scales as the viscosity of the bath fluid $\eta_{bath}$. }
\end{figure*}

\subsection{Rheometry}
We use a stress-controlled rotational rheometer (MCR 92, Anton Paar) for rheological characterization of the Carbopol suspension bath fluid. To minimize wall slip, a sandblasted $50\;\si{\milli\meter}$ diameter plate-plate geometry is used at a gap size of $500\;\si{\micro\meter}$. Each experiment starts with a preshear in both positive and negative directions performed at a shear rate $\dot{\gamma}=300\,\si{\per\s}$ for $60\;\si{\second}$ each, followed by an equilibration for $180\;\si{\second}$ . The steady-state viscosity is measured in the shear rate range of $\dot{\gamma}=0.01 - 100\;\si{\per\second}$ first by gradually lowering $\dot{\gamma}$ from $100$ to $0.01\;\si{\per\second}$, and then increasing it back up to $100\;\si{\per\second}$. To obtain the storage modulus $G'$ and the loss modulus $G''$, a frequency sweep is performed at a strain amplitude $\gamma_0=0.005$ and an amplitude sweep at a frequency $\omega=6.28\;\si{\radian\per\second}$. Both experiments are conducted to ensure frequency independence and linearity of the moduli. For the characterization of the transient viscoelasticity during recovery, we first run a shear start-up step at a constant shear rate $\dot{\gamma}=0.01, 0.1, 1.0\;\si{\per\second}$ up to the total strain of $\gamma=1.0$, at which the stress $\sigma_0$ is set to zero. The strain response during the resulting fluid-to-solid transition is used to estimate the transient elastic modulus and the transient viscosities. All experiments are conducted at room temperature $T=21\si{\celsius}$. \par

%Results: I. Spreading of low-viscosity ink filaments

\section{Results}
\subsection{Dependence of penetration length on bath concentration and speed}
The penetration length $l$ depends on both the microgel concentration $c$ of the bath fluid and the bath translation speed $V_{bath}$, as shown in Fig.~\ref{pics_heatmap}(a,b). At lower concentrations, the penetration length $l$ increases with $c$, while for higher $c$, $l$ moderately decreases. At a given concentration, $l$ decreases nearly monotonically with the bath speed $V_{bath}$, although its dependence on $V_{bath}$ is less pronounced than that on $c$. We attribute this trend to the varying viscoelasticity and the yield stress of Carbopol suspensions in different concentration regimes. No spreading is observed when $V_{bath}=0\,\si{\milli\meter\per\s}$, which indicates that this phenomenon is not caused by the wetting of the needle by the ink. \par

%Results: II. Viscous pressure drop in shear-thinning fluid baths
\subsection{Viscous pressure drop in shear-thinning fluid baths}
To elucidate the ink spreading at lower bath concentrations, where the shear-thinning fluid exhibits no solidlike elasticity and yield stress, we first note that the spreading is observed even in a bath of glycerol, a Newtonian fluid, as shown in Fig.~\ref{p_drop_N}(a). Such upward flow of the ink is not observed when the orientation of the needle is parallel to the direction of bath displacement \cite{Trikalitis2023}, although the dispensed ink far from the needle slowly but continuously rises due to buoyancy, hence slanting the filament, as displayed in Fig.~\ref{p_drop_N}(b). This absence of strong vertical ink flow right behind the needle suggests that the pressure right below the tip ($P_1$) is approximately equal to the pressure right above the tip ($P_2$) as expected from the symmetry of the flow. By contrast, the asymmetry between the flow above and below the tip when the needle is perpendicular to the bath displacement causes $P_2$ to be lower than $P_1$, as illustrated in Fig.~\ref{p_drop_N}(a), leading to the vertical flow of the ink. \par

\begin{figure}[b]
\setlength{\abovecaptionskip}{-3pt}
\hspace*{0cm}\includegraphics[scale=0.5]{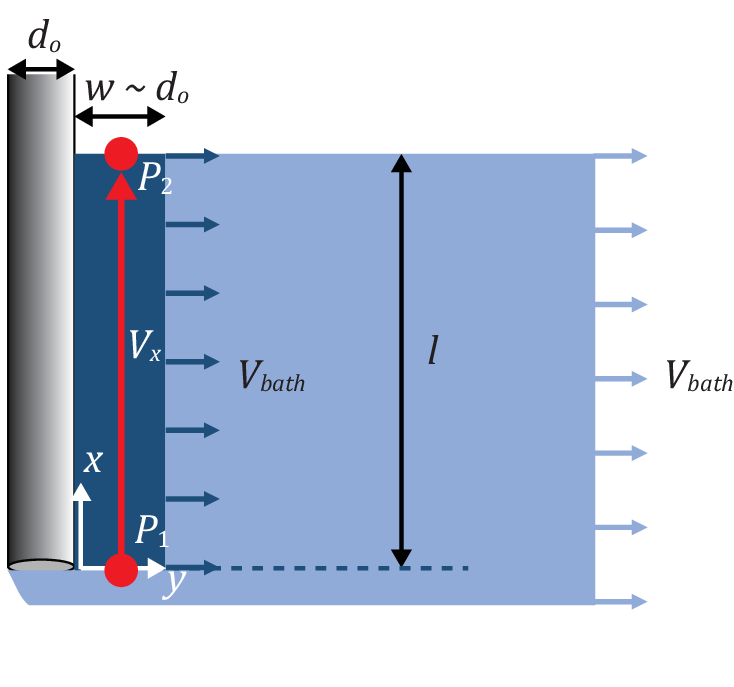}
\caption{\label{sp_model} Schematic illustrating the spreading ink flow due to viscous pressure drop across the needle. The vertical flow is assumed to occur within a column (dark blue) of width $w\,\sim\,d_o$ in both $y$ and $z$ directions, where $d_o$ denotes the outer diameter of the needle. As the bath moves at a constant speed $V_{bath}$, a uniform outflow at $V_{bath}$ occurs across the penetration length $l$.}
\end{figure}

\begin{figure*}[t]
\setlength{\abovecaptionskip}{-30pt}
\hspace*{0cm}\includegraphics[scale=0.20]{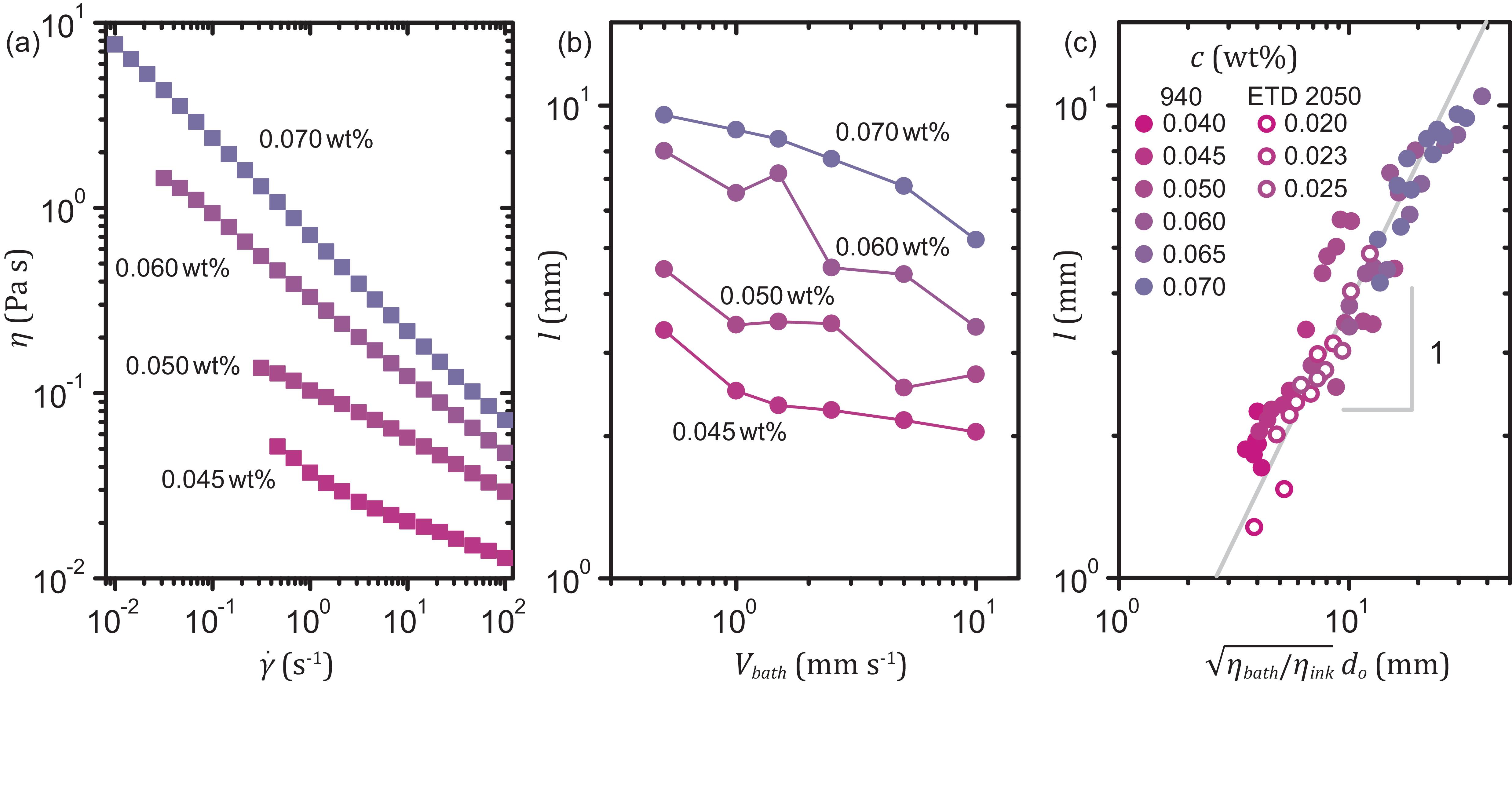}
\caption{\label{low_c_scaling} (a) Apparent viscosity $\eta$ of Carbopol 940 suspensions as a function of the shear rate $\dot{\gamma}$ for different particle concentrations $c$, where no yield stress is observed. (b) Penetration length $l$ as a function of the bath speed $V_{bath}$ for the corresponding $c$. (c) Linear scaling relation between $l$ at various bath speeds $V_{bath}$ and the square root of the viscosity ratio $\sqrt{\eta_{bath}/\eta_{ink}}$ for low-$c$ Carbopol 940 and ETD 2050 suspensions, whose yield stresses are zero.}
\end{figure*}

The pressure gradient in the vertical direction in the Newtonian fluid bath results from the viscous pressure drop across the needle. For fluid elements traveling around the cylinder, the friction from the outer surface dissipates mechanical energy through viscosity, which leads to a pressure drop downstream. A scaling relation derived from the Stokes equation $\boldsymbol{0}=-\boldsymbol{\nabla} p+\eta_{bath}\nabla^2\boldsymbol{v_{bath}}$ dictates that the pressure drop
\begin{equation}
P_{\infty}-P_2\;\sim\;\frac{\eta_{bath}V_{bath}}{d_o}, \label{p_drop}
\end{equation}
where $p$ denotes the pressure field, $\eta_{bath}$ the bath fluid viscosity, $\boldsymbol{v_{bath}}$ the velocity field, $P_{\infty}$ the far-field pressure and $d_o$ the outer diameter of the needle. For fluid elements traveling below the needle, however, the local velocity field remains nearly the same as the upstream uniform flow, which suggests that $P_{1}\approx P_{\infty}$. The pressure decrease $\Delta P\equiv P_{1}-P_{2}$ in the upward direction behind the needle can therefore be expressed as  
\begin{equation}
\Delta P\;\sim\;\frac{\eta_{bath}V_{bath}}{d_o}, \label{p_diff}
\end{equation}
as illustrated in Fig.~\ref{p_drop_N}(c). The pressure difference due to gravity between the two points is neglected in this relation, since its effect is canceled out by the hydrostatic pressure gradient in the surrounding fluid. \par

When this viscous pressure drop in the Newtonian bath fluid generates the vertical flow of the ink, the penetration length $l$ scales as the square root of the bath-to-ink viscosity ratio. Given the pressure set by the surrounding bath fluid, a column of the dispensed ink fluid forms right behind the needle, whose end pressures are equal to $P_1$ and $P_2$, as shown in Fig.~\ref{sp_model}. Assuming that both horizontal dimensions of the ink column are of the order of the needle outer diameter $d_o$, we derive a scaling relation from the steady-state momentum equation along the needle:
\begin{equation}
\frac{\Delta P}{l}\;\sim\;\frac{\eta_{ink}V_x}{{d_{o}}^2}, \label{momen_scaling}
\end{equation}
where $V_x$ denotes the characteristic ink velocity in the needle direction. Since the bath translates at a constant speed $V_{bath}$ driving ink out of the column in which the upward flow occurs, the mass conservation holds when
\begin{equation}
V_{x}{d_{o}}^2 \;\sim\; V_{bath}ld_o. \label{m_conserv}
\end{equation}
Rearranging Eqs.~\eqref{p_diff}\eqref{momen_scaling}, and \eqref{m_conserv}, yields
\begin{equation}
l\;\sim\;\sqrt{\frac{\eta_{bath}}{\eta_{ink}}}d_o, \label{pen_scaling}
\end{equation}
which reflects that the spreading is driven by the bath fluid viscosity $\eta_{bath}$ but is resisted by the ink fluid viscosity $\eta_{ink}$. \par

\begin{figure}[t]
\setlength{\abovecaptionskip}{-20pt}
\hspace*{-0.10cm}\includegraphics[scale=0.20]{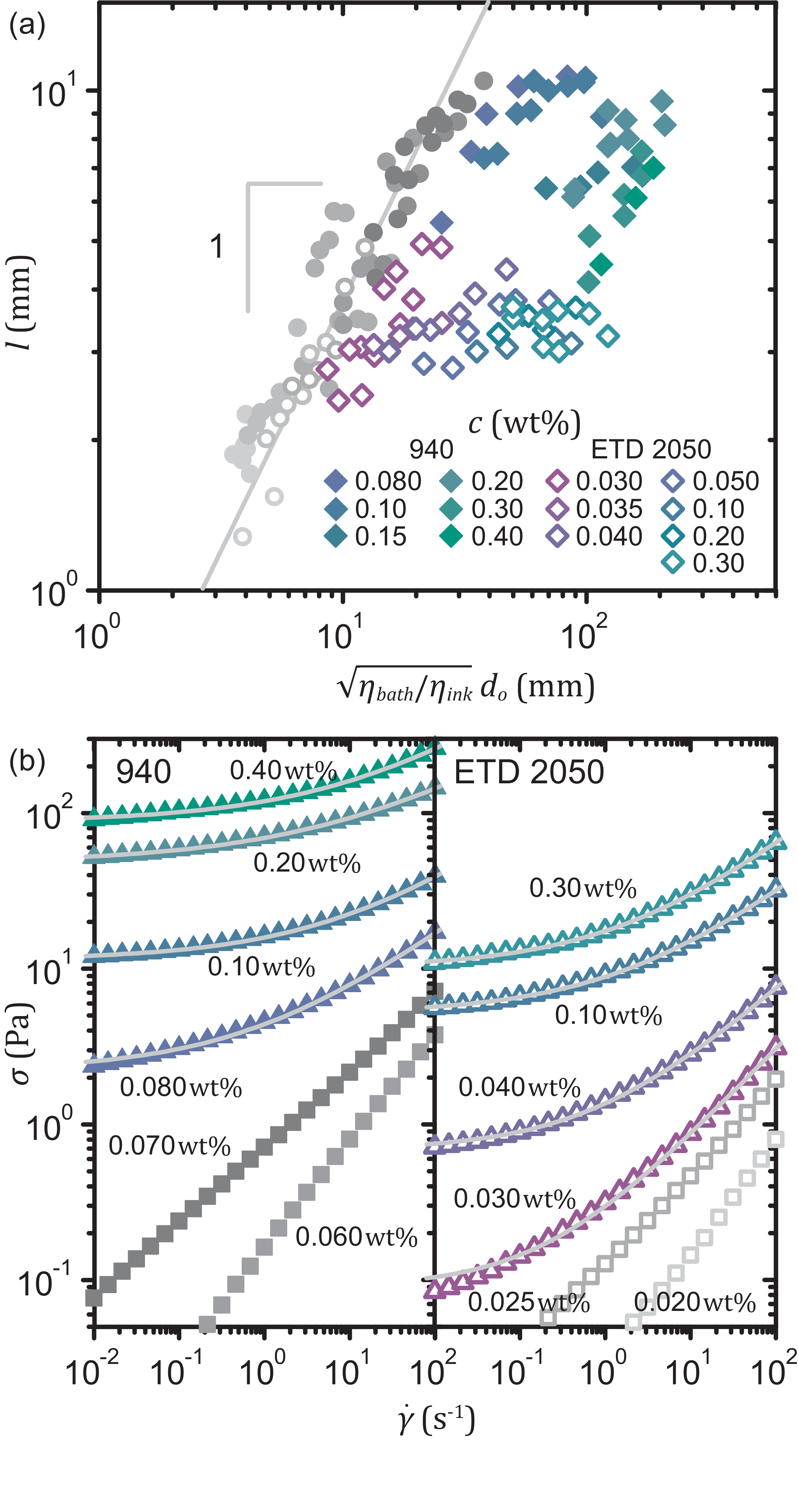}
\caption{\label{yield_stress} (a) Deviation from the linear scaling relation between $l$ and $\sqrt{\eta_{bath}/\eta_{ink}}d_{0}$ for higher-$c$ Carbopol suspensions (colored). (b) Flow curves of the higher Carbopol concentrations (colored) that exhibit nonzero yield stresses. Grey lines represent the Herschel-Bulkley model fitted to the data for each concentration. The flow curves at lower concentrations (grey squares) do not form plateaus at low shear rates $\dot{\gamma}$.}
\end{figure}

This competition between the opposing effects of the viscosities of the two fluids accounts for the ink spreading in the shear-thinning Carbopol suspensions. The viscosity $\eta$ of Carbopol suspensions increases with the microgel concentration $c$ and decreases with the shear rate $\dot{\gamma}$, as shown in Fig.~\ref{low_c_scaling}(a), when $c$ is kept sufficiently small such that no yield stress is observed. Thus, given the same ink fluid, it can be inferred from Eq.~\eqref{pen_scaling} that the penetration length $l$ increases with $c$ and decreases with the bath speed $V_{bath}$, as verified by the experimental results displayed in Fig.~\ref{low_c_scaling}(b) and previously in Fig.~\ref{pics_heatmap}(a). Evaluating the right-hand side of Eq.~\eqref{pen_scaling}, using the values of $\eta_{bath}$ corresponding to the effective shear rates ${\dot{\gamma}}_{e}\equiv V_{bath}/{d_{o}}$ indeed leads to a linear relationship between the measured penetration length $l$ and $\sqrt{\eta_{bath}/\eta_{ink}}d_o$ for both types of Carbopol particles, as shown in Fig.~\ref{low_c_scaling}(c), which validates the application of the Newtonian fluid model to the flow behavior in the shear-thinning fluids. \par

\begin{figure*}[t]
\setlength{\abovecaptionskip}{-20pt}
\hspace*{-0.0cm}\includegraphics[scale=0.20]{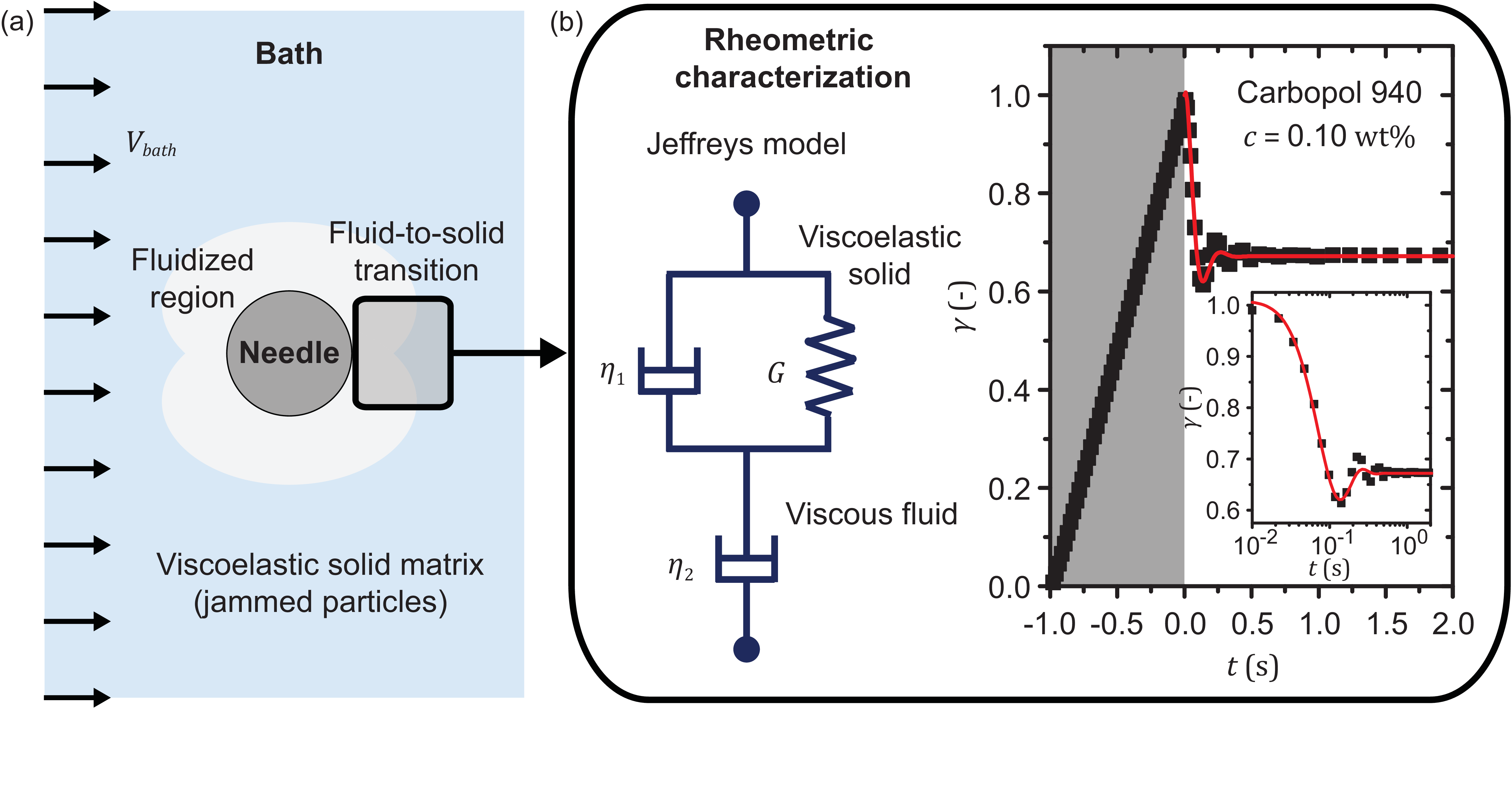}
\caption{\label{const_recov} (a) Top-view schematic of a Carbopol bath around the needle. The yield stress fluid is locally fluidized upstream right in front of the needle and solidified downstream right behind the needle, forming a fluidized region (white) through which the ink can spread. (b) Rheometric characterization via constrained recovery experiments that mimic the fluid-to-solid transition behind the needle. The Jeffreys model, a 3-parameter viscoelastic analog model, is fitted to the strain response $\gamma(t)$ when the stress is set to zero, as shown by the red line. Inset: $\gamma(t)$ plotted on the log timescale.}
\end{figure*}

%Results: III. Viscoelastic pressure drop in yield stress fluid baths
\subsection{Viscoelastic pressure drop in yield stress fluid baths}
In bath fluids of higher microgel concentrations, where nonzero yield stress is observed, the ink spreading cannot be solely attributed to the pressure drop across the needle due to the bath fluid viscosity. The penetration length $l$ does not increase linearly with $\sqrt{\eta_{bath}/\eta_{ink}}d_o$ at the higher concentrations, as shown in Fig.~\ref{yield_stress}(a). For either Carbopol 940 or Carbopol ETD 2050, this deviation from the scaling in Eq.~\eqref{pen_scaling} occurs at the concentrations where the corresponding flow curves can be approximated by the Herschel-Bulkley model $\sigma=\sigma_{y}+K{\dot{\gamma}}^{n}$, where $\sigma$ denotes the shear stress, $\sigma_{y}$ the yield stress, $K$ the consistency index, and $n$ the power-law exponent \cite{Divoux2011}, as displayed in Fig.~\ref{yield_stress}(b). A nonzero value of the yield stress $\sigma_{y}$ signifies jamming of the microgel particles in the bath, which gives rise to a solidlike elastic stress field due to the deformation caused by the translating needle \cite{Oelschlaeger2022}. We therefore hypothesize that the ink spreading in the yield stress fluid stems from a pressure gradient due to both the viscosity and the elasticity of the Carbopol suspensions. \par

The ink spreading due to the viscoelasticity of the yield stress fluid warrants a characterization of its transient rheological behavior, rather than its steady-state behavior, as the bath fluid constantly undergoes a fluid-to-solid transition behind the needle. During the translation, the bath fluid is locally yielded and fluidized in a confined region around the needle \cite{Putz2008,Tokpavi2008,Tokpavi2009,Grosskopf2018}. The bath fluid then quickly recovers the yield stress in the areas where the needle has passed, as illustrated in the schematic of Fig.~\ref{const_recov}(a). Hence, any bath fluid element traveling adjacent to the needle undergoes a change of state from fluid to solid in the region behind the needle, where the ink column forms. \par

We examine the viscoelasticity of the bath fluid during its fluid-to-solid transition by performing an analogous experiment on the rheometer. In each experiment, the Carbopol suspension is sheared at a constant shear rate $\dot{\gamma}$ up to a total strain $\gamma=1.0$, at which the sample is released by applying zero stress $\sigma=0\,\si{\pascal}$. The total strain value of $\gamma=1.0$ is chosen based on the assumption that the maximum shear strain observed in the fluidized region during printing is of the order of $1$. However, we confirm that similar results are obtained even with a much larger strain $\gamma=10$. The resolidification behind the needle is simulated by the subsequent constrained recovery of the strain $\gamma$ while the stress is set to zero, a commonly used method to probe transient viscoelastic properties of various yield stress fluids \cite{Singh2021}. We also find that the results depend only marginally on the initial shear rate within the range of $\dot{\gamma}=0.01-1.0\,\;\si{\per\second}$, and thus simply use the data obtained at $\dot{\gamma}=1.0\,\;\si{\per\second}$ for each bath fluid in the following analysis. \par

We parameterize the viscoelastic behavior during the constrained recovery by fitting the transient response of the Jeffreys model, derived in Appendix~\ref{Jeff_params}, to the strain as a function of time. A linear, three-parameter analog model that constitutes a serial connection of a viscoelastic solid part and a viscous fluid part \cite{Baravian1998,Ewoldt2007,BenmouffokBenbelkacem2010}, the Jeffreys model serves as a reasonable approximation of the temporal change in the strain, as illustrated in Fig.~\ref{const_recov}(b). The Jeffreys model is a simplest model that can mimic a partial, delayed strain recovery with viscoelastic oscillations \cite{Ewoldt2007}, and has been used to describe a transient behavior of Carbopol suspensions \cite{BenmouffokBenbelkacem2010}. The values of the two viscosities $\eta_1$ and $\eta_2$ and the elastic modulus $G$, obtained from the curve fits, reveal both similarities and differences between the steady-state and the transient rheology of Carbopol suspensions, as described in Appendix~\ref{trans_params}. \par

To account for the effects of the bath fluid viscoelasticity on the ink spreading, we estimate the penetration length induced by the viscous pressure drop and that by the elastic deformation separately. By replacing the steady-state viscosity $\eta_{bath}$ in Eq.~\eqref{pen_scaling} with the transient viscosity $\eta_{2}$ of the viscous fluid part of the Jeffreys model, we expect the penetration length caused by the viscous pressure drop across the needle to be
\begin{equation}
l_v\;\sim\;\sqrt{\frac{\eta_{2}}{\eta_{ink}}}d_{o}. \label{viscous_part}
\end{equation}
We speculate that the pressure difference along the ink column due to the elastic stress applied by the surrounding solid matrix is of the order of
\begin{equation}
\Delta P\;\sim\; G\gamma_{c} \;\sim\; G, \label{Pdrop_elastic}
\end{equation}
where the characteristic shear strain $\gamma_{c}$ is assumed to be 1, given a single geometric length scale $d_{o}$ in the plane perpendicular to the needle. When combined with Eqs.~\eqref{momen_scaling} and \eqref{m_conserv}, Eq.~\eqref{Pdrop_elastic} yields the following scaling relation for the penetration length caused by the bath fluid elasticity:
\begin{equation}
l_{e}\;\sim\;\sqrt{\frac{Gd_{o}}{\eta_{ink}V_{bath}}}d_{o}. \label{elastic_part}
\end{equation}
Although Eq.~\eqref{viscous_part} is independent of the bath speed $V_{bath}$ because the transient viscosity $\eta_{2}$ is independent of the shear rate, the elastic contribution to the spreading expressed in Eq.~\eqref{elastic_part} incorporates the $V_{bath}$ dependence of the penetration length $l$. \par

\begin{figure}[t]
\setlength{\abovecaptionskip}{-30pt}
\hspace*{-0.02cm}\includegraphics[scale=0.20]{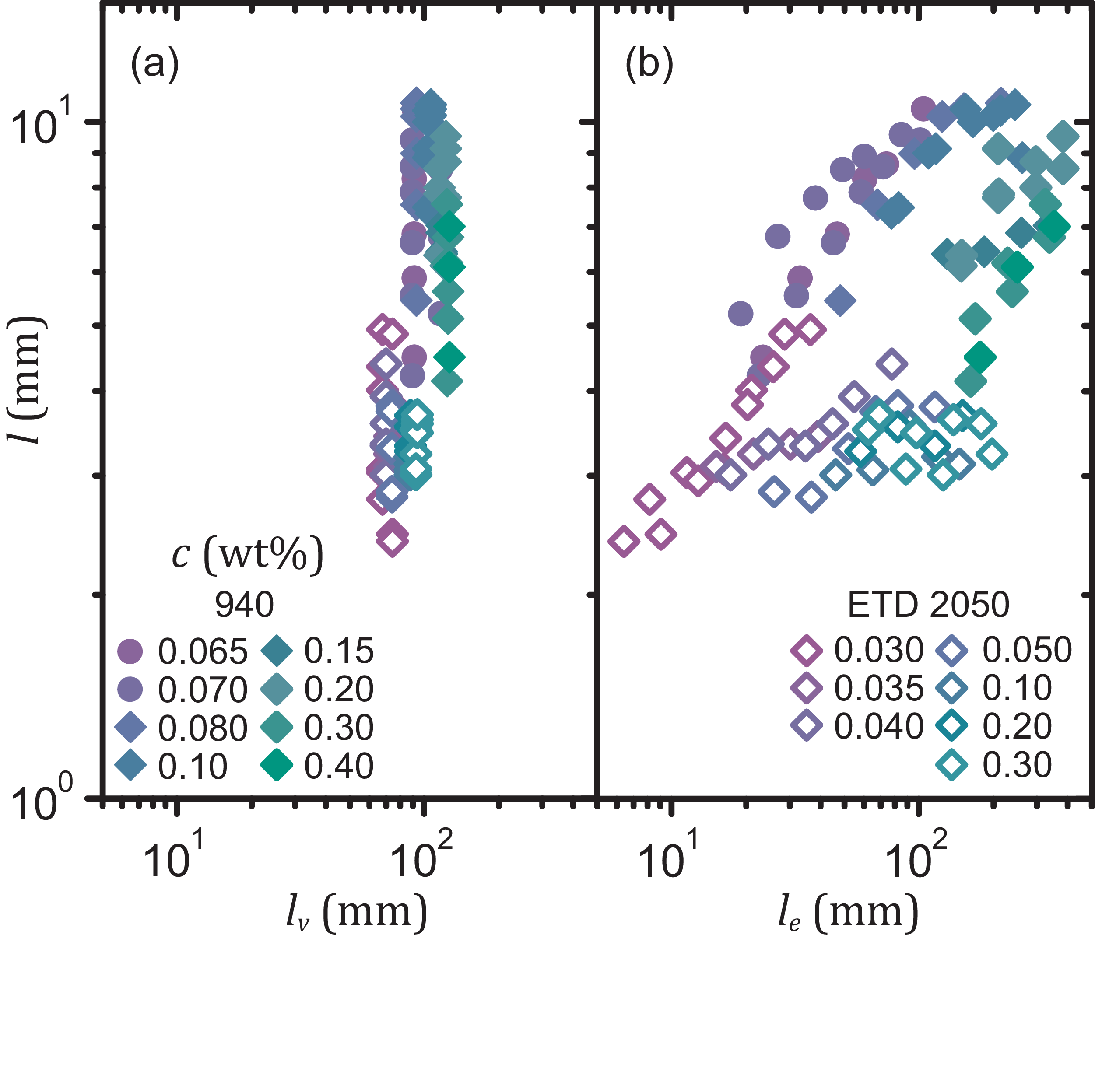}
\caption{\label{viscoelastic_scaling} Penetration length $l$ at various bath speeds $V_{bath}$ as a function of (a) the viscous penetration length scale $l_v$ and (b) the elastic penetration length scale $l_e$ for high Carbopol concentrations.}
\end{figure}

The resulting estimates of the viscous and the elastic penetration lengths, $l_{v}$ and $l_{e}$ respectively, indicate that the pressure difference due to the elasticity is indeed comparable to the viscous pressure drop for the high Carbopol concentrations at or above the jamming transition. The values of $l_{v}$ and $l_{e}$ calculated with the transient viscoelastic parameters are of the same order as displayed in Fig.~\ref{viscoelastic_scaling}(a,b). As $l_{e}$ increases more rapidly with the concentration $c$ than $l_{v}$ for either Carbopol type, $l_{e}$ even exceeds $l_{v}$ at the highest concentrations. Neither dominates the other within the concentration ranges investigated in this work, however, which obscures how exactly the measured penetration length $l$ depends on $l_{v}$ and $l_{e}$. \par

Greater than the corresponding penetration length $l$ approximately by an order of magnitude, the values of $l_{v}$ and $l_{e}$ also hint that the spreading is hindered at the high Carbopol concentrations despite the combined effects of viscoelasticity. We ascribe the hindrance to the yield stress. When the yield stress relative to the viscous stress, as quantified by the Oldroyd number $Od\,\equiv\,\sigma_{y}/\left(K{\dot{\gamma}}^{n}\right)$ based on the Herschel-Bulkley model parameters, increases, the volume of the region of fluidization around the needle is known to decrease \cite{DeBesses2003,Tokpavi2008,Fraggedakis2016,Grosskopf2018,Hewitt2018}. The reduced volume of the fluidized Carbopol suspension would suppress the elongation of the ink column, hence decreasing the penetration length. We find that the Oldroyd number $Od=\sigma_{y}/\left(K{\dot{\gamma}_{e}}^{n}\right)$ is of order $1$ for most of the Carbopol suspensions with nonzero yield stresses, as shown in Appendix~\ref{Od}, suggesting nonnegligible influence of the yield stress on the local flow around the needle. The interplay between the yield stress and the viscoelastic forces in determining the penetration length remains to be explored. \par

\begin{figure}[t]
\setlength{\abovecaptionskip}{10pt}
\hspace*{0cm}\includegraphics[scale=0.5]{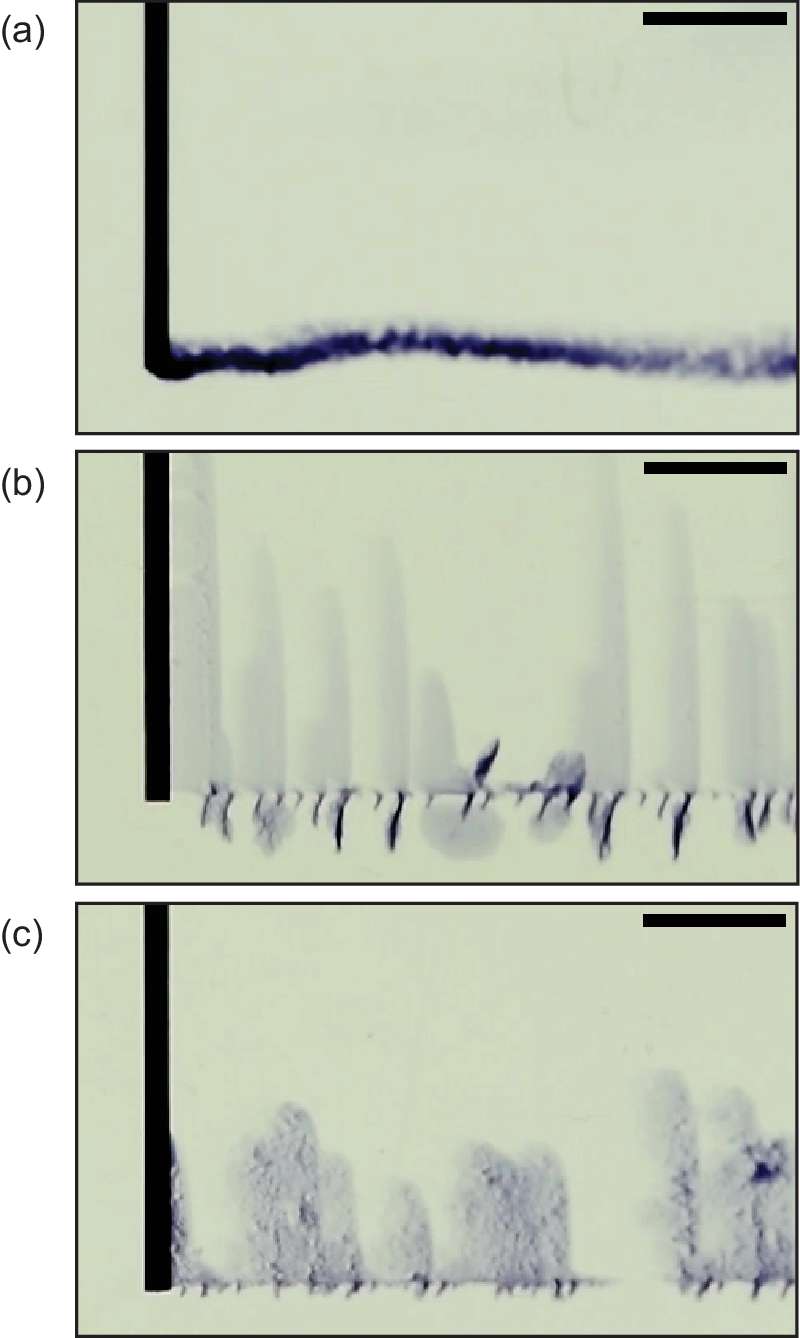}
\caption{\label{ext_conc} Structural defects of deposited filaments at extreme Carbopol concentrations. (a) Bent filament in a dilute Carbopol 940 suspension ($c=0.03\,\si{\wtpercent}$), and traces left by the intermittent bursts of the injected ink in (b) a dense Carbopol 940 suspension ($c=0.30\,\si{\wtpercent}$ ), and (c) a dense Carbopol ETD 2050 suspension ($c=0.30\,\si{\wtpercent}$), all at a bath speed $V_{bath}=0.5\,\si{\milli\meter\per\s}$. Each scale bar denotes $5\,\si{\milli\meter}$.}
\end{figure}

%Results: IV. Defects at low and high microgel concentrations
\subsection{Defects at low and high microgel concentrations}
Though our results indicate that the penetration length $l$ is reduced when the concentration is much lower or much higher than the critical concentration at jamming for either type of Carbopol, printing the low-viscosity ink in bath fluids at such extreme concentrations is susceptible to the formation of other structural defects. At very low Carbopol concentrations, the viscosity of the suspension is of the same order of magnitude as the viscosity of water ($\sim1\;\si{\milli\pascal\s}$) as the microgel particles minimally interact with one another under shear. The inertia of the ink fluid injected into such a low-viscosity bath is not negligible even for the lowest injection speed $V_{ink}=0.5\;\si{\milli\meter\per\s}$ in our experiments. The corresponding Reynolds number $Re\equiv\rho V_{ink} d_{i}/\eta_{bath}\approx0.5$, where $\rho=1000\;\si{\kilo\gram\per\cubic\meter}$ denotes the density of water, is of order 1. Furthermore, given $V_{ink}=V_{bath}$, $Re\simeq1$ indicates that the vorticity generated within the bath fluid around the translating needle can be advected over a considerable distance without dissipation, thus deforming the deposited ink filament. The structure of the printed ink filament is indeed unstable, as shown in Fig.~\ref{ext_conc}(a).
\par

At high Carbopol concentrations, the ink is injected intermittently, as displayed in Fig.~\ref{ext_conc}(b,c). Although the finger-like shape of the spread ink may be reminiscent of the Saffman-Taylor instability observed in quasi-2D geometries \cite{Saffman1958,Homsy1987}, this phenomenon differs from the Saffman-Taylor instability in that the structures emerge from intermittent injection rather than an unstable translating interface. Also, this defect cannot be ascribed to any fracture or cavity formation in the Carbopol suspension, as no evidence of cavities within the bath is observed. \par

The unsteadiness of the ink flow arises from the yield stress of the bath fluid that results in a stress buildup within the injection system via the radial expansion of the soft tubing that connects the syringe and the injection needle. For such high concentrations, the yield stress $\sigma_y$ of the Carbopol suspension is significantly high, such that the portion of the bath fluid underneath the tip of the moving needle does not continuously fluidize \cite{Hutchens2016,Mohammadigoushki2023}. In fact, the yield stress $\sigma_y$ for the concentration $c=0.10\;\si{\wtpercent}$ at which the intermittent flow starts to be observed, as shown in Fig.~\ref{pics_heatmap}(b) for Carbopol 940, is approximately $10\;\si{\pascal}$, sufficiently large to induce considerable expansion of the tubing during the continuous injection of the ink from the syringe. The increase in the radius $\delta_r$ due to an internal pressure increment $\Delta p$ can be evaluated using the theory of linear elasticity for thin-walled tubes as $\delta_r={\Delta}pR^{2}/(hE)$, where $R$ denotes the original tube radius, $h$ the wall thickness, and $E$ the Young's modulus. This radial expansion would increase the cross-sectional area by $2\pi R\delta_r=2\pi{\Delta}pR^3/(hE)$, which requires the ink to be injected into the tubing (but not yet into the bath) for the lag time ${\Delta}t\approx2\pi{\Delta}pR^3L/(hEQ_{ink})$, where $L$ denotes the tubing length, $Q_{ink}$ the volumetric flow rate. We estimate the distance traveled by the needle during the lag time to be 
\begin{equation}
V_{bath}{\Delta}t=8{\Delta}pR^3L/(hE{d_i}^2)\approx0.90\,\si{\milli\meter}, \label{tubing_expansion}
\end{equation}
when ${\Delta}p=\sigma_y\approx10\;\si{\pascal}$, comparable to the needle outer diameter $d_{o}=0.908\,\si{\milli\meter}$. For our polyvinyl chloride tubing, $R\approx1.99\,\si{\milli\meter}$, $L\approx1.0\,\si{\meter}$, $h=0.794\,\si{\milli\meter}$, and $E\approx2.4\,\si{\mega\pascal}$. At the high concentrations, it is therefore energetically more favorable for the pressurized ink to radially expand the tubing than to be injected into the bath, until the pressure inside the tubing increases by the yield stress, which takes longer time than the one required for the needle to traverse its own diameter. Increasing the bath speed $V_{bath}$, and thus the injection speed $V_{ink}=V_{bath}$, tends to suppress the bursts, as displayed in Fig.~\ref{pics_heatmap}(b), which suggests that the higher shear stress applied by the more rapidly translating needle to the surrounding bath fluid facilitates yielding of the Carbopol suspension around the needle tip. The minimum bath speed required for a stable ink flow increases with the concentration, as the yield stress increases. \par

%Discussion and conclusions
\section{Discussion and conclusions}

By dispensing water threads in baths of Carbopol suspensions, we show that the ink spreading, a unique type of defect commonly observed in embedded printing, results from the pressure gradient in the needle direction set by the viscoelasticity of the bath fluid. For low concentration Carbopol suspensions that exhibit shear thinning without yielding, the penetration length of the ink behind the needle scales as the square root of the bath-to-ink viscosity ratio at the corresponding effective shear rate, because of the pressure field generated by the viscous flow around the needle. For high concentration Carbopol suspensions that behave like soft solids due to their nonzero yield stresses, the stress around the needle due to the elastic modulus is comparable to the viscous pressure difference. Hence, the dependence of the penetration length on the Carbopol concentration and the bath speed can be better understood as the combined effects of the elasticity and the viscosity. The local fluidization and resolidification of the bath fluid around the needle for high Carbopol concentrations warrant a characterization of the viscoelasticity during the fluid-to-solid transition. We parameterize the transient viscoelasticity by fitting the Jeffreys model to the strain response in constrained recovery experiments. We find that the viscous and elastic pressure gradients estimated by the transient viscoelastic parameters are of comparable scales, and hence neither viscosity nor elasticity alone fully accounts for the penetration length at and above the jamming transition. The values of the penetration length 
for the high Carbopol concentrations, however, are an order of magnitude lower than the expected length scales based on the viscoelasticity, which may be attributed to the reduced area of fluidization behind the needle due to the high yield stress. \par

These results suggest that the spreading of the low-viscosity ink can be best mitigated by utilizing baths at either very low or very high Carbopol concentrations. Yet the baths at very low concentrations fail to stabilize the dispensed ink as they cannot effectively dissipate the fluid kinetic energy, while the baths at very high concentrations suffer unsteady ink flow caused by the competition between the high yield stress and the elastic stress in the elastomer injection tubing. The presence of these distinct artifacts highlights that there are optimal levels of the viscosity and the yield stress of the bath fluid at which the deposition of the low-viscosity ink is best controlled in embedded printing. The challenge in preparing the optimal bath lies in that the viscosity and the yield stress (and the elasticity) cannot be independently tuned for most complex fluids. \par

A deeper understanding of the interactions between the ink and the bath fluids may be obtained by the flow visualization around the needle and the incorporation of extra rheological parameters in the model. Our model in this work accounts for only the ink flow in the vertical direction behind the needle. Experimental characterization of the ink flow at the needle tip upon its injection would add another dimension to the model, which may elucidate the role of the shear applied by the surrounding bath fluid or the mechanism for the intermittent bursts at higher Carbopol concentrations. A 3D visualization of the bath fluid would support or refute the hypothesis on the reduction of the fluidized area for higher Carbopol concentrations. The translation of the needle through the bath may induce local extensional flows behind the needle, in which the extensional viscosity of the bath fluid may come into play \cite{Nelson2019,Ewoldt2022}. Even the part of the bath fluid in shear flow may alter the pressure gradient in the ink column via its normal stress due to its non-Newtonian nature \cite{deCagny2019}. Our findings provide a good starting point for the development of more advanced models through such investigations.   \par

\begin{figure*}[t]
\setlength{\abovecaptionskip}{-30pt}
\hspace*{-0.15cm}\includegraphics[scale=0.20]{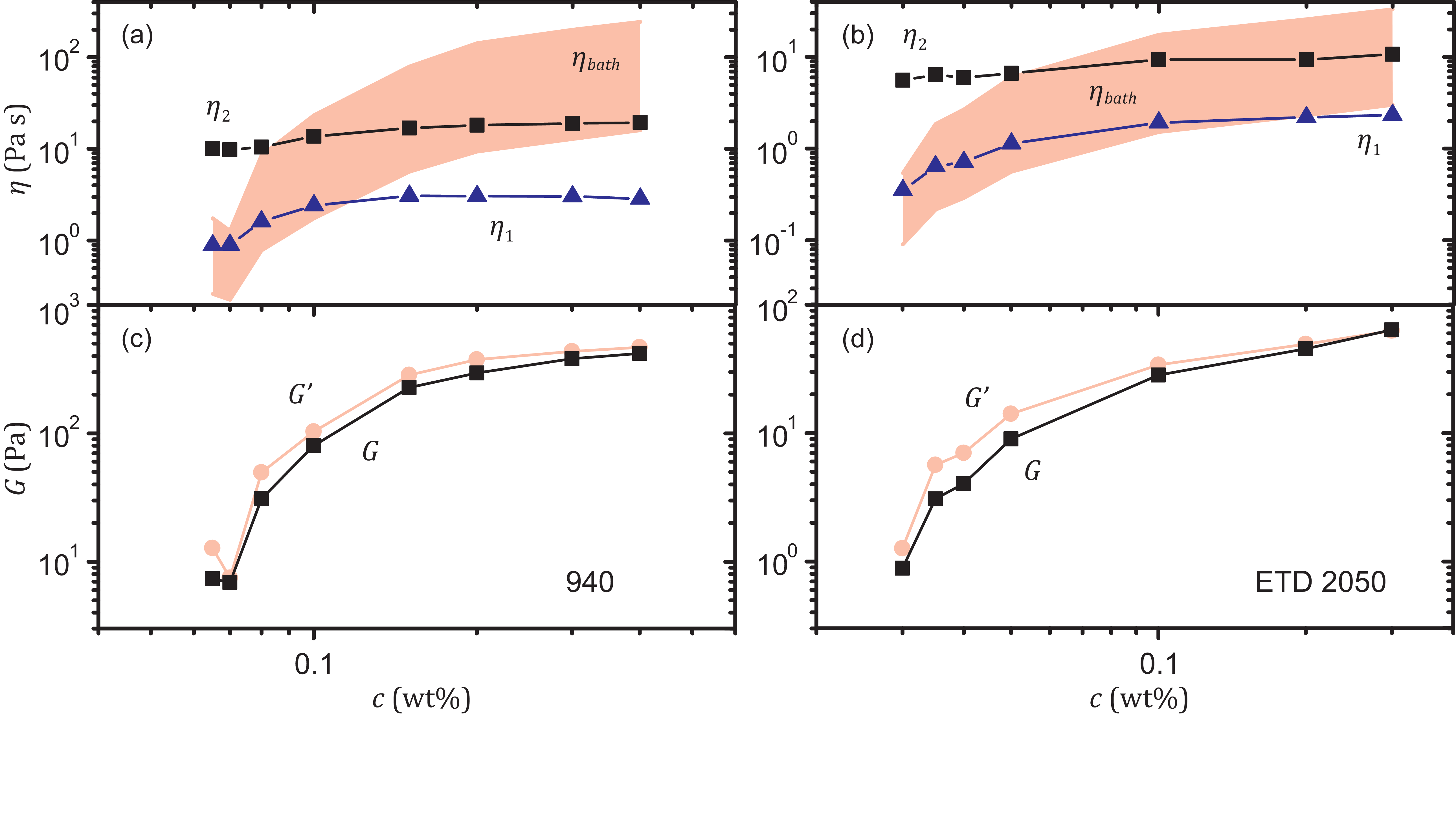}
\caption{\label{trans_params_fig} (a,b) Comparison of the transient Jeffreys viscosities $\eta_1$, $\eta_2$ and the steady-state apparent viscosity $\eta_{bath}$ and (c,d) that of the transient Jeffreys elastic modulus $G$ and the storage modulus $G'$ at a frequency $\omega=6.28\,\si{\radian\per\s}$ for Carbopol 940 (a,c) and Carbopol ETD 2050 (b,d) suspensions at different concentrations. The shaded areas in (a) and (b) represent the ranges of $\eta_{bath}$ for the shear rate $\dot{\gamma}=0.55-11\,\si{\per\s}$, which corresponds to the range of the effective shear rates $\dot{\gamma}_{e}=V_{bath}/d_{o}$ for the bath speeds ($0.5-10\,\si{\milli\meter\per\s}$) used in this work.}
\end{figure*}

\begin{acknowledgments}
We acknowledge support from the University of California Cancer Research Coordinating Committee. We thank Lubrizol for generously providing the Carbopol samples.
\end{acknowledgments}

%Appendices
\appendix

\section{Calculation of Jeffreys parameters from constrained recovery experiments}
\label{Jeff_params}
The functional form of the strain $\gamma(t)$ of the Jeffreys model in constrained recovery can be derived by directly solving for the strain term in the momentum equation for the rotating body of the stress-controlled rheometer and the stress-strain constitutive equation for the model \cite{Baravian1998,Ewoldt2007,BenmouffokBenbelkacem2010}. In our experiments, in which the rheometer applies zero stress for $t\geq0$, Newton's second law applied to the combined mass of the instrument and the upper plate dictates that
\begin{equation}
\frac{I}{b}\ddot{\gamma}(t)=\left[1-H(t)\right]\sigma_{0}-\sigma_{s}(t), \label{Newtons_law}
\end{equation} 
where $I$ denotes the rotational inertia of the instrument and the geometry, $b$ the geometry factor that converts the angular displacement and the torque to the strain and the stress, respectively, $H(t)$ the Heaviside function, $\sigma_0$ the steady-state stress applied by the rheometer during the flow at $t<0$, and $\sigma_s$ the stress applied to the sample. Note that $\sigma_s$ is not necessarily zero for $t>0$, as the instrument accelerates. For a plate-plate geometry, $b={\pi}R^4/(2h)$, where $R$ is the plate radius and $h$ the gap size \cite{Ewoldt2007}. The constitutive relation for the Jeffreys model can be expressed as
\begin{equation}
\left(\eta_{1}+\eta_{2}\right)\dot{\sigma_{s}}(t)+G\sigma_{s}(t)=\eta_{2}G\dot{\gamma}(t)+\eta_{1}\eta_{2}\ddot{\gamma}(t), \label{J_constitutive}
\end{equation}
where $\eta_1$, $\eta_2$, and $G$ denote the model parameters, as shown in the schematic of Fig.~\ref{const_recov}(b). \par

Substituting the expression for $\ddot{\gamma}$ from Eq.~\eqref{Newtons_law} into Eq.~\eqref{J_constitutive} and differentiating with respect to time leads to 
\begin{equation}
\begin{split}
&\left(\eta_1+\eta_2\right)\ddot{\sigma}_{s}+\left(G+\frac{\eta_{1}\eta_{2}b}{I}\right)\dot{\sigma}_{s}+\frac{\eta_{2}Gb}{I}\sigma_{s} \\
&\qquad=-\frac{\eta_{2}Gb}{I}\sigma_{0}H(t)-\frac{\eta_{1}\eta_{2}b}{I}\sigma_{0}\delta(t)+\frac{\eta_{2}Gb}{I}\sigma_{0}, \label{sig_ode}
\end{split}
\end{equation}
where $\delta$ denotes the Dirac delta function. By solving for the sample stress $\sigma_{s}(t\geq0)$ in the Laplace domain with a set of initial conditions $\sigma_{s}(t=0^{-})=\sigma_0$ and $\dot{\sigma}(t=0^{-})=0$, it can be expressed as
\begin{equation}
\sigma_{s}(t)=\sigma_{0}\exp(-Bt)\left[\cos(At)+C\sin(At)\right], \label{sig_s}
\end{equation}
where
\begin{align}
A&=\frac{\sqrt{-\left(GI-\eta_{1}\eta_{2}b\right)^{2}+4GI{\eta_{2}}^{2}b}}{2\left(\eta_{1}+\eta_{2}\right)I} \label{paramA} \\
B&= \frac{GI+\eta_{1}\eta_{2}b}{2\left(\eta_{1}+\eta_{2}\right)I} \label{paramB} \\
C&=\frac{GI-\eta_{1}\eta_{2}b}{\sqrt{-\left(GI-\eta_{1}\eta_{2}b\right)^{2}+4GI{\eta_{2}}^{2}b}}. \label{paramC}
\end{align}
Substituting Eq.~\eqref{sig_s} into Eq.~\eqref{Newtons_law} and integrating twice for $t>0$ yields
\begin{equation}
\begin{split}
&\gamma(t)=-\frac{b\sigma_0}{I(A^2+B^2)^2}\exp(-Bt)\left[X_1\cos(At)-X_2\sin(At)\right] \\
&\qquad\quad +\frac{b}{I}K_{1}t+\frac{b}{I}K_{2}, \label{gamma_fit}
\end{split}
\end{equation}
where
\begin{align}
X_1&=B^2-A^2+2ABC \label{paramX1} \\
X_2&=A^2C-B^2C+2AB \label{paramX2},
\end{align}
and $K_1$ and $K_2$ denote constants of integration. Since the strain has to reach a constant value as $t\rightarrow\infty$, $K_{1}$ is assumed to be zero. After calculating the value of $K_2$ from the late-time plateau of the strain $\gamma(t)$ in the experiments, we fit a curve of the functional form in Eq.~\eqref{gamma_fit} to the experimental $\gamma(t)$ while imposing the initial condition $\gamma(t=0)=1.0$ to ensure continuity of the strain. Using Eqs.~\eqref{paramA}-\eqref{paramX2}, we calculate the Jeffreys parameters $\eta_1$, $\eta_2$, and $G$, as well as the steady-state stress $\sigma_{0}$. We confirm the validity of the fitting results by comparing the value of $\sigma_{0}$ to that of the actual stress applied just prior to the application of zero stress. \par

\begin{figure}[t]
\setlength{\abovecaptionskip}{-30pt}
\hspace*{-0.15cm}\includegraphics[scale=0.20]{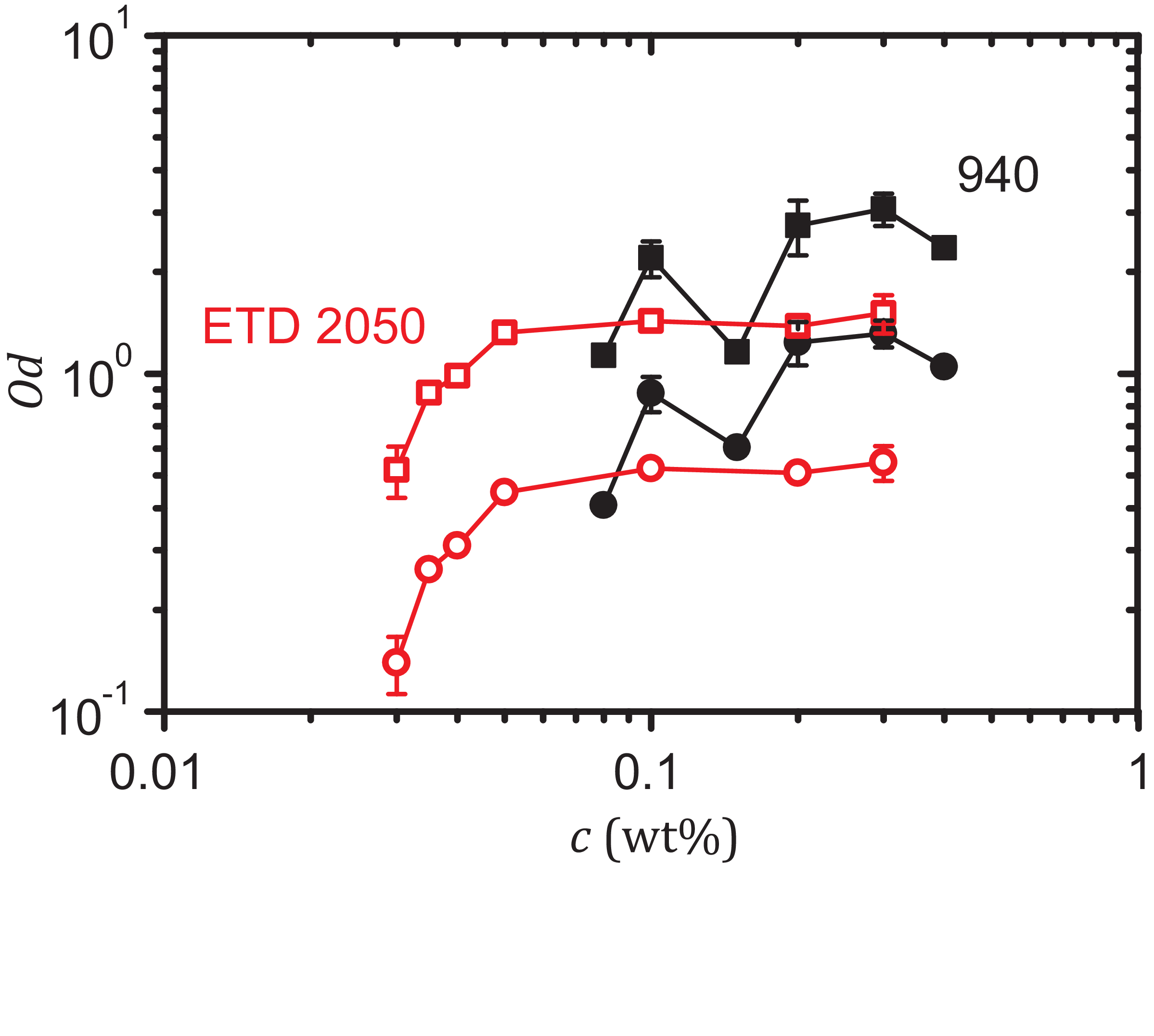}
\caption{\label{Od_fig} Oldroyd number $Od$ at the effective shear rates $\dot{\gamma}_{e}=V_{bath}/d_o$ corresponding to $V_{bath}=1.0\,\si{\milli\meter\per\s}$ (squares) and $V_{bath}=10.0\,\si{\milli\meter\per\s}$ (circles) as a function of the Carbopol concentration $c$ for Carbopol 940 (black filled symbols) and Carbopol ETD 2050 (red open symbols).}
\end{figure}

\section{Comparison of transient and steady-state rheological parameters}
\label{trans_params}

Although both transient viscosities $\eta_1$ and $\eta_2$ tend to increase with the microgel concentration $c$ for either Carbopol type, their concentration dependence is much weaker than that of the steady-state viscosity $\eta_{bath}$, as displayed in Fig.~\ref{trans_params_fig}(a,b). This pronounced difference may arise from that the steady-state viscosity originates in constant shearing and resultant deformation of particles out of equilibrium, whereas the transient viscosities during the recovery characterize the transition back into equilibrium upon the removal of the shearing force. The transient elastic modulus $G$, by contrast, closely follows the steady-state storage modulus $G'$ for all the concentrations of either Carbopol type, as shown in Fig.~\ref{trans_params_fig}(c,d), which may likewise be attributed to the fact that the system remains close to mechanical equilibrium during the small-amplitude oscillatory shear. The storage modulus $G'$ is nearly independent of the frequency $\omega$ for all the Carbopol suspensions.

\section{Dependence of Oldroyd number $Od$ on bath speed $V_{bath}$ and concentration $c$}
\label{Od}

The Oldroyd number $Od=\sigma_{y}/(K{\dot{\gamma}_{e}}^n)$ is mostly of order 1 within the range of bath speeds $V_{bath}$ studied in this work, as shown in Fig.~\ref{Od_fig}. In addition, $Od$ increases with the concentration $c$ for both types of Carbopol, which may illuminate why the penetration length $l$ tends to decrease with $c$ for the highest concentrations. \par

\bibliography{ink_spreading.bib}

\end{document}